\documentclass[aps,prl,twocolumn]{revtex4-2}

\usepackage[english]{babel}

\usepackage{amsmath}
\usepackage{graphicx}
\usepackage{xcolor}

\usepackage{ulem}

\begin{document}
\title{Force on a sphere suspended in flowing granulate}
\author{{Jing Wang}$^1$, {Bo Fan}$^{2,5}$, {Tivadar Pong\'o}$^{3,4}$, {Tam\'as B\"orzs\"onyi}$^2$, {Ra\'ul Cruz Hidalgo}$^3$ and {Ralf Stannarius}$^1$}

\affiliation{$^1$Institute of Physics, Otto von Guericke University, Magdeburg, Germany\\
$^2$Institute for Solid State Physics and Optics, Wigner Research Centre for Physics, Budapest, Hungary\\
$^3$F\'isica y Matem\'atica Aplicada, Facultad de Ciencias, Universidad de Navarra,
Pamplona, Spain\\
$^4$Collective Dynamics Lab, Division of Natural and Applied Sciences, Duke Kunshan University, Kunshan, China\\
$^5$Physical Chemistry and Soft Matter, Wageningen University $\&$ Research, Wageningen, The Netherlands
}

\date{\today}

\begin{abstract}
We investigate the force of flowing granular material on an obstacle. A sphere suspended in a discharging silo experiences both the weight of the overlaying layers and drag of the surrounding moving grains.
In experiments with frictional hard glass beads, the force on the obstacle was practically flow-rate independent. In contrast, flow of nearly frictionless soft hydrogel spheres added drag to the gravitational force.
The dependence of the total force on the obstacle diameter is qualitatively different for the two types of material: It grows quadratically with the obstacle diameter in the soft, low friction material, while it grows much weaker, nearly linearly with the obstacle diameter, in the bed of glass spheres.
In addition to the drag, the obstacle embedded in flowing low-friction soft particles experiences a total force from the top as if immersed in a hydrostatic pressure profile, but a much lower counterforce acting from below. In contrast, when embedded in frictional, hard particles, a strong pressure gradient forms near the upper obstacle surface.

\end{abstract}

\maketitle

What happens to an object immersed in flowing granular material?
In common fluids, drag and buoyancy forces on objects are well-understood and textbook knowledge.
In granular matter, however, the situation is much more complicated. Already at the bottom of static granular heaps, forces can have counter-intuitive characteristics that depend upon the deposition procedure
\cite{vanelPRE1999,Matuttis1998,Loevoll1999,Ai2013}. The pressure at the bottom of silos containing hard grains does not grow linearly with fill height, but saturates at a certain level, as observed already in 19$^{th}$ century by Hagen \cite{Hagen1852,Tighe2010} and Janssen \cite{Janssen1895,Sperl2006}.
When an intruder moves relative to the granulate, the situation is even more complex.
An important research field (e.~g. \cite{Ambroso2005,Joubaud2014,Katsuragi2007,Katsuragi2013,Katsuragi2014,Huang2020,Clark2012,Clark2013,Clark2014,Clark2016,vanderMeer2017}) is
the impact of objects into granular beds. Both velocity and penetration depth determine the related drag forces.
In quasi-twodimensional (2D) beds of photoelastic disks, force chains can be visualized  \cite{Clark2012,Clark2013,Clark2014,Clark2016}. A review of impact experiments was given by van der Meer \cite{vanderMeer2017}.

In the geometry considered by Pacheco-V\'azquez et al.~\cite{Pacheco2009}, the impact was horizontal.
Force measurements on horizontally moving objects basically decouple gravitation and drag.
In such experiments, the objects are typically
immersed in continuously moving  granular beds, or they are pulled horizontally in a bed at rest \cite{Wieghardt1952,Wieghardt1975,Albert1999,Albert2001,Albert2001a,Brzinski2010,Costantino2011,Artoni2019,Kobayakawa2018,Hilton2013,Takada2020}. For slow motion, the drag force is  independent of velocity and scales linearly with the cross-section of the intruder. Horizontal drag may cause uplift forces \cite{Ding2011,Guillard2014}. The immersion 
of the grains in liquids
can mimic variable
gravity effects when the liquid density is varied \cite{Costantino2011}.
There is evidence that vibrations may affect the velocity dependence of drag forces \cite{Seguin2017}.
For a spherical intruder in granular shear flow,
Jing et al.~\cite{Jing2022} reported analogies to Stokes drag.

Other typical geometries are spheres, rods or horizontal plates lifted in a grain bed
\cite{hillEPL2005,Zhou2004,Hossain2020a}, or plunged in a granulate at constant vertical speed \cite{hillEPL2005,stonePRE2004,Brzinski2013,Kang2018,Roth2021}. In those studies, the drag force was velocity independent but increased nearly linearly with penetration depth. Numerical studies also included intruders in fluidized beds and quicksand \cite{Zaidi2020,Zaidi2020a}.
In soft particle beds, a constant creep may be reached, as shown by Dijksman and Mullin \cite{Dijksman2022}.

Experiments measuring drag force on objects (spheres, cylinders, circular plates or cones) suspended in an emptying silo \cite{tsunakawaPT1975,atkinsonCES1983,Chehata2003,moyseyPT2013,kobylkaPT2019}, that were confirmed in DEM simulations \cite{moyseyPT2013,kobylkaPT2019,coppinCPM2022}, found a velocity-independent force.
In these cases, the grain--grain contacts are mobilized everywhere in the
flowing material, unlike the above mentioned cases with intruders
moving in a static granular bed. In  silos or pipes,
sometimes objects are placed inside with the
aim of controlling the flow field. There, it is important to know
the force exerted on such objects by the flowing granulate. A highly relevant situation is a vertically suspended object in downward flowing grains, like in silo discharge.

\begin{figure}[htbp]
\centering
\includegraphics[width=0.6\columnwidth]
{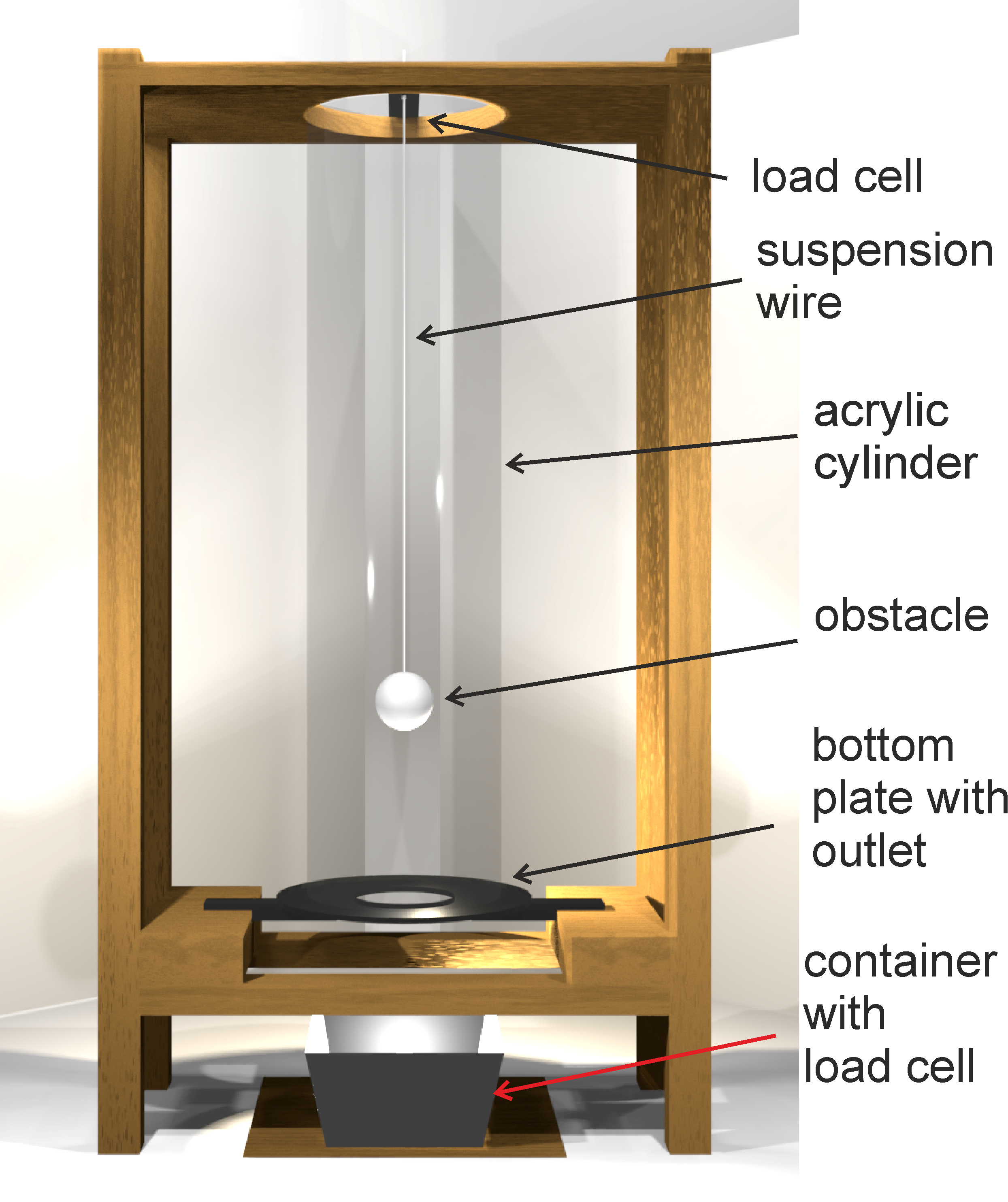}
\caption{Schematic drawing of the setup.}
\label{fig:setup}
\end{figure}

We analyze forces on spherical obstacles in a 3D cylindrical silo. The geometry is sketched in Fig.~\ref{fig:setup}. A thin wire holds a sphere in the center of the silo at a fixed height of 20 cm (1.4 times the silo diameter $D_{\rm silo} = 14.4$ cm)
above the bottom, where in absence of the obstacle, the downward flow would be sufficiently uniform.
The wire is fixed to a load cell that measures the vertical force.
The vertical flow velocity was adjusted by the outlet diameter $D_{\rm 0}$ at the silo bottom.
Two types of grains were compared: low-friction, soft hydrogel spheres (HGS) \cite{Ashour2017h,Harth2020} with 7.5~mm diameter and an elastic modulus of the order of 50~kPa, and hard glass spheres (GLS) with 3.15~mm diameter. The obstacles had diameters $d$ between 10~mm and 40 mm, their masses ranged from 0.83~g to 53~g.
The corresponding weights were negligible with respect to  forces exerted by the surrounding grains.
Nevertheless, we corrected all measured data for the obstacle weights.

In addition, we ran discrete element modeling (DEM) simulations, computed the relevant continuous fields, and performed a micromechanical analysis. In addition, we ran discrete element modeling (DEM) simulations, computed the relevant continuous fields, and performed a micromechanical analysis. Details of the DEM implementation \cite{rubio2017large} based on a Hertz-Mindlin model \cite{Poeschel2005} and post-processing coarse-graining implementation \cite{goldhirsch2010stress} are provided as supplemental material \cite{Supplement}.

\begin{figure}[htbp]
\centering
\includegraphics[width=\columnwidth]
{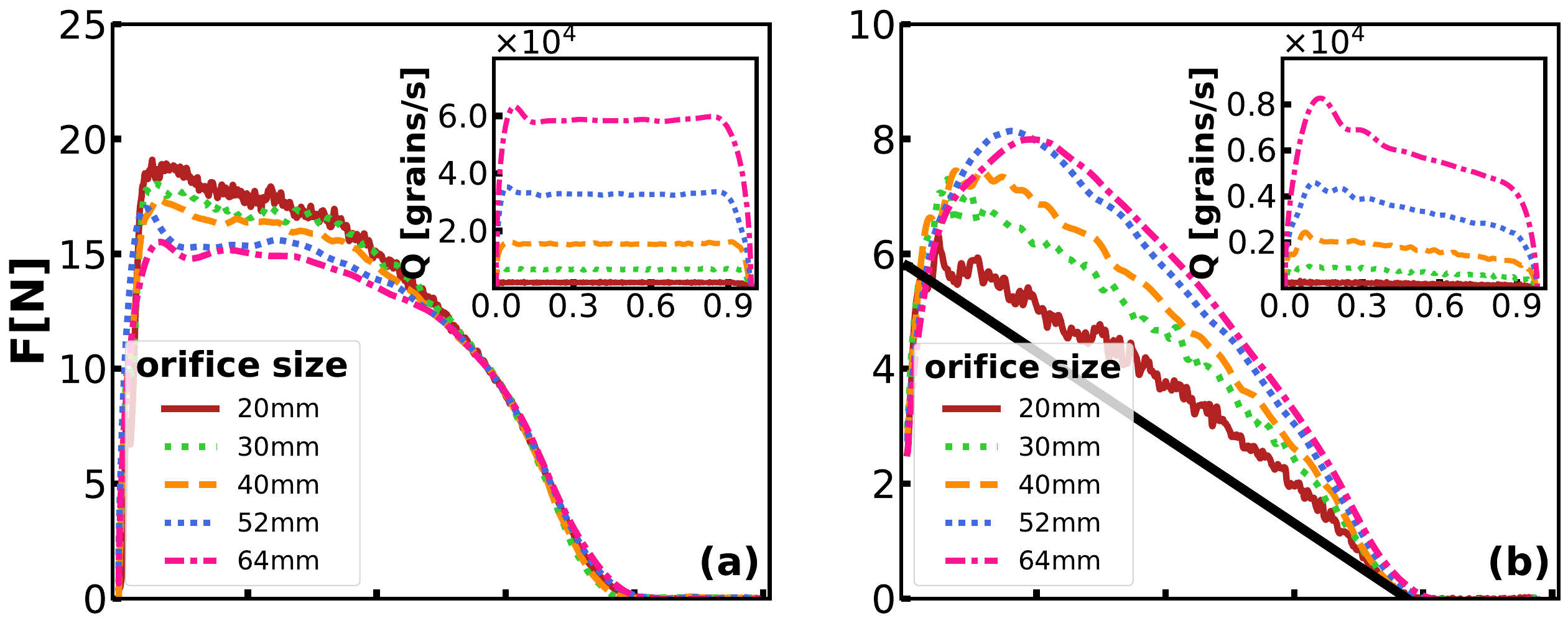}
\includegraphics[width=\columnwidth]
{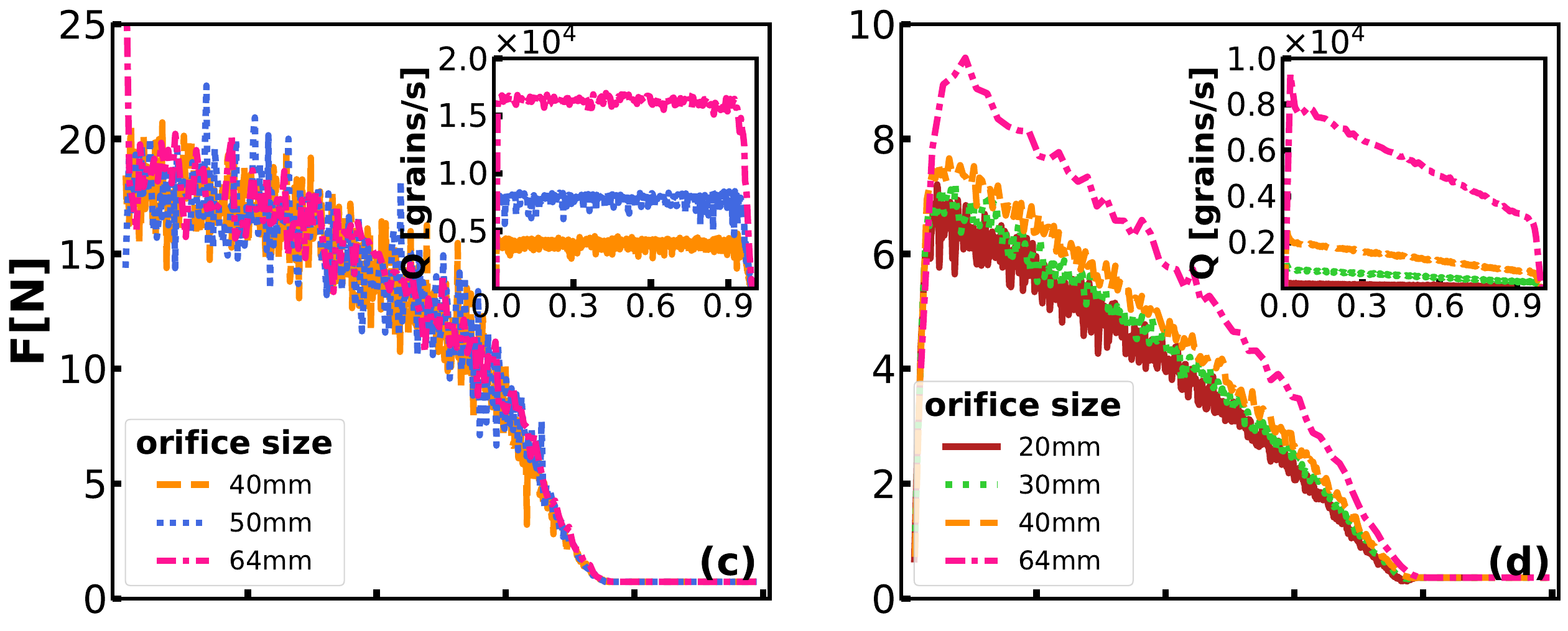}
\includegraphics[width=\columnwidth]{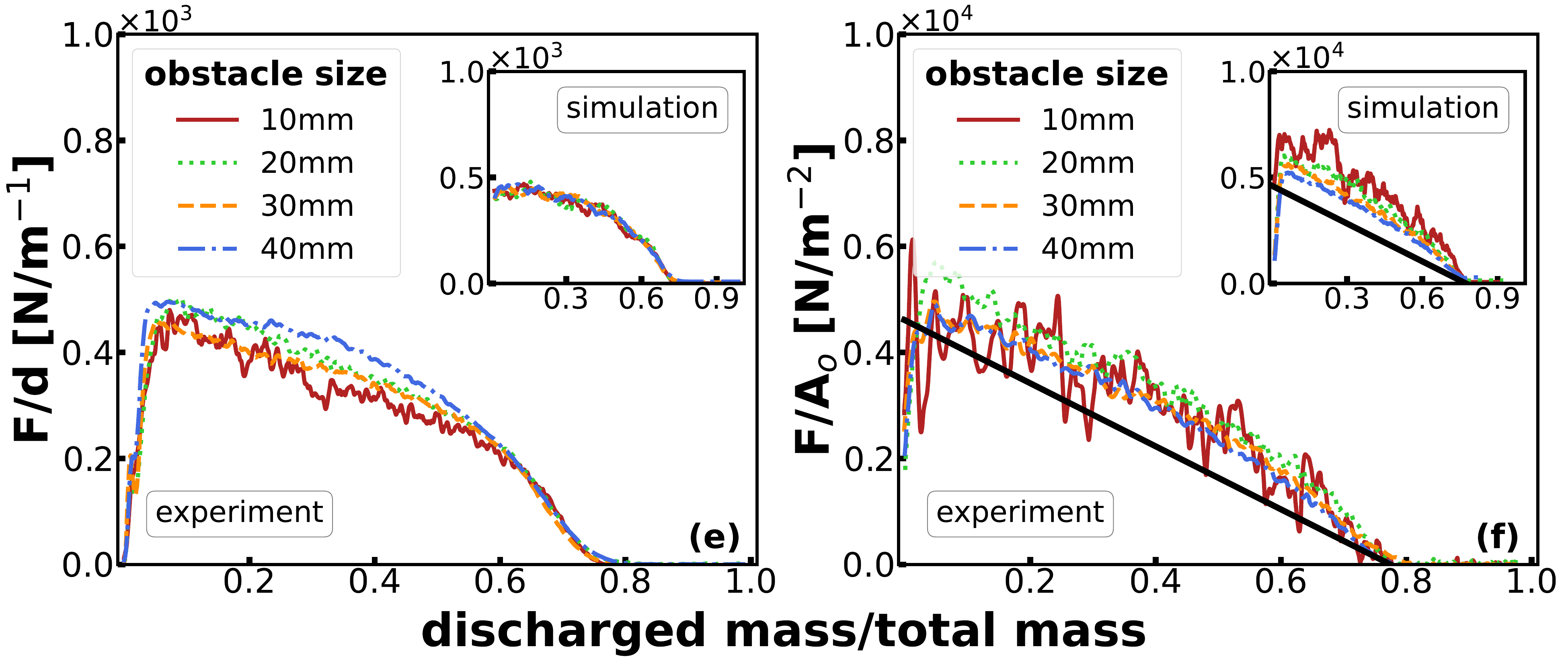}
\caption{
Force acting on an obstacle with 40 mm diameter submerged in GLS (a) experiments, (c) numerical, and in HGS (b) experiments, (d) numerical, plotted as a function of the discharged mass in terms of the total mass. The insets show the outflow rates $Q$ for the respective orifice sizes, same abscissae as subfigures.
Panel (e) illustrates the experimental force obtained for GLS rescaled by the obstacle diameter (inset: numerical data); and panel (f) shows the experimental force obtained for HGS rescaled by the cross-sectional area (inset: numerical data). Different obstacle sizes are compared for fixed $D_0=20$~mm.
Black lines in (b) and (f) indicate the weight of the material in a hypothetical cylindrical column above the obstacle.
}
\label{fig:sketch5}
\end{figure}

The forces on the balls consist of two contributions, those arising from the weight of grains above,  mediated by force chains, and those related to friction with the flowing grains. With different outlet sizes $D_0$ we controlled the flow speed, whereby the flow profiles remained unchanged at the position of the obstacle.
Figures \ref{fig:sketch5}(a,c) show the results for the hard GLS material, obtained experimentally and numerically. Except for short transients, flow rates for given orifice sizes
are constant (see insets).
In the range of orifices chosen, the flow rate $Q$ varies
by a factor of $\approx 30$ \footnote{For GLS, a flow rate $Q=10^4$ grains/s in the experiment leads to $\approx 1.5$ cm/s flow at the obstacle position. For HGS,
$Q = 10^3$ grains/s corresponds to slightly less than 2 cm/s flow speed.
}.
The forces acting on the obstacle in
Fig.~\ref{fig:sketch5}(a),
however, show only a slight trend of reduced force with larger flow speed.
In repeated experiments, this trend was not always clearly present. Mostly, the differences were within the experimental uncertainty. In the numerical simulations (see Fig. \ref{fig:sketch5}(c)), the force variations with discharge rate were minimal.

Since the largest flow rate ($D_0=64$~mm) is about 30 times larger than the smallest rate ($D_0=20$~mm), we assume that the force measured with the smallest orifice is pretty close to a `quasi-static' value expected in absence of drag forces (which should not be mistaken for the force on a sphere in a static bed, which depends sensitively on the filling procedure and history). The static character of the major force contributions was confirmed experimentally for a 30 mm wide orifice and obstacles of 20, 30 and 40 mm diameter: When we interrupted the outflow while the silo was still filled to 30 cm height, the force on the ball remained practically unchanged \cite{Supplement}.

\begin{figure}[t]
\centering
\includegraphics[width=0.66 \columnwidth]
{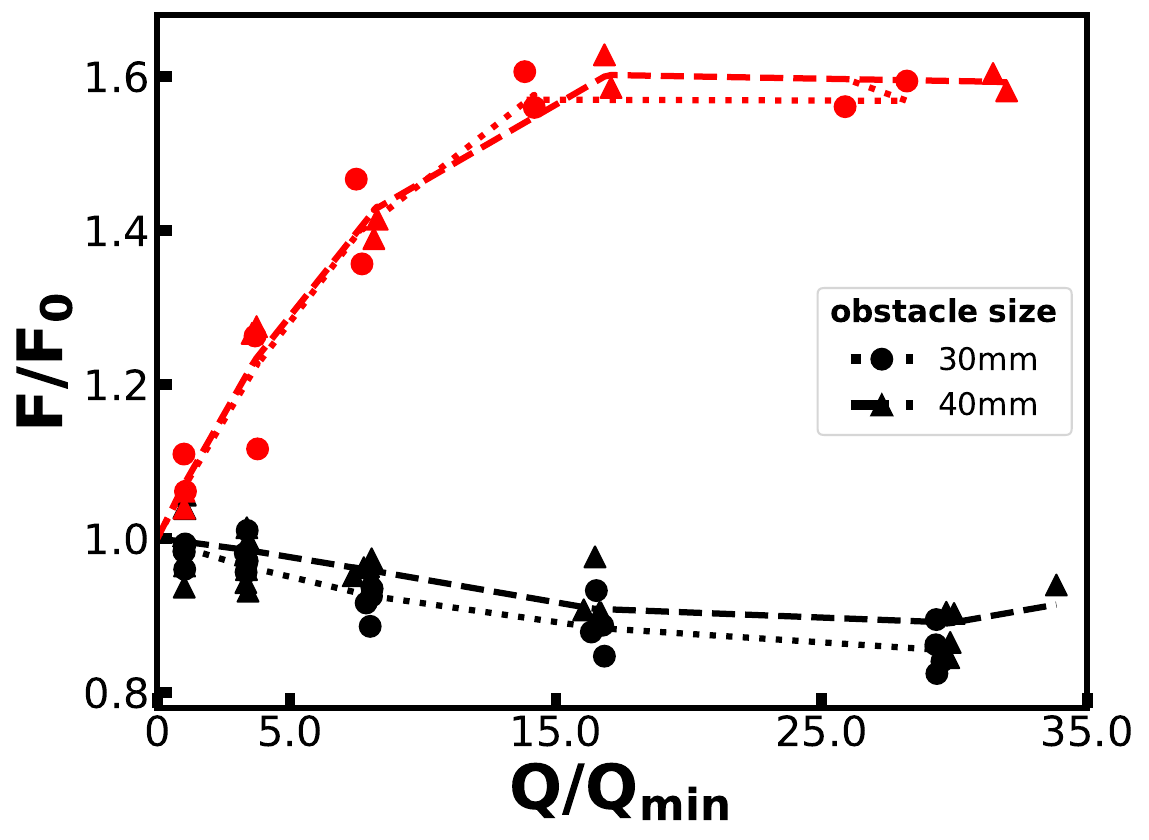}
\caption{Effect of the flow on the forces experienced by the obstacle: Flow rates are scaled with the lowest value measured for the $D_0=20$~mm orifice. Forces are scaled such that the curves can be extrapolated to 1 for $Q_{\min}$. 
Black symbols refer to GLS, red ones to HGS. Lines guide the eye.}
\label{fig:forceflow}
\end{figure}

Figures~\ref{fig:sketch5}(b,d) show
the results for the soft HGS material.
In contrast to hard grains, the discharge rates of HGS are known to be fill level dependent \cite{pongoNJP2021,Harth2020}.
This was confirmed in our present measurements (see insets in Figs.~\ref{fig:sketch5}(b,d)).
The forces on the obstacle in HGS are clearly velocity dependent. This time, remarkably, larger flow speeds led to increasing forces.
Force measurements after an abruptly stopped flow are consistent with
this: After the outflow of HGS was interrupted, the forces on the obstacle dropped substantially \cite{Supplement}.
A sudden decay after the outflow ceased was followed by a slow further reduction, until a plateau was reached representing a residual static pressure.
Figure~\ref{fig:forceflow} shows the discharge rates $Q$ and the maximum forces $F$ for two obstacle sizes ($d=30$~mm and $40$~mm). Discharge rates were
scaled with the rate for the $D_0=20$~mm orifice, and
forces were scaled with an approximate extrapolated value $F_{0}$ for vanishing flow. 
This plot clearly evidences the opposite effects of flow in both types of materials. Apparently, a saturation is reached for sufficiently large flow rates in both cases. For small orifice sizes ($<50$~mm) we are 
in the slow flow regime \cite{Albert1999} $v<\sqrt{2g d}/10$, whereas for the largest orifices, the critical slow flow limit is reached. Possible explanations for the observed saturation are different for the HGS and GLS material. For hard spheres flowing around a suspended cylinder in quasi-2D, Chehata et al.~\cite{Chehata2003} reported an independence of drag forces on flow speeds at comparable obstacle and grain sizes. Their velocities were much higher, and thus the saturation found here is consistent with their observations.
For the soft grains, the saturation has a simpler reason: For the largest orifice sizes, the flow becomes so fast that the local packing density lowers, small cavities may form. This reduces the drag force and compensates drag effects of larger velocities.

Comparing the forces on balls of different diameters, another striking difference between soft, low friction HGS and normal friction hard GLS is evident, as documented in Figs.~\ref{fig:sketch5}(e-f): An obstacle suspended in soft, slippery grains experiences a force that is in good approximation proportional to its diameter squared. 
It is noticeable in Fig. \ref{fig:sketch5}(f) that all the plots coincide within the experimental and numerical accuracy, while the squared radius varies by a factor of 16.
This can be understood even in a quantitative manner: In a static HGS column, the pressure $p$ grows nearly linearly, quasi-hydrostatically, with the depth $h$ beneath the granular bed surface \cite{Ashour2017h}, with $p \approx h\cdot 7$~kN/m.
This is consistent with an initial total force of $\approx 85$~N measured at the silo bottom. The redirection of forces to the container walls is ineffective because of the low friction coefficient. For an initial 80~cm fill level, the pressure on the ball at 20~cm height is thus 4.2~kPa.
 The black line in Fig.~\ref{fig:sketch5}(f) indicates how the pressure would change when one assumed in first approximation a hydrostatic pressure and a linear drop of the fill level with discharged mass. When the fill level reaches the obstacle, the pressure vanishes. The pressure curve multiplied by the projected area, viz. the horizontal cross-section area of the suspended ball, yields the corresponding force on the obstacle's top surface. For the $d=40$~mm ball, one finds initially 5.3~N, and a trend indicated by the black line in Fig.~\ref{fig:sketch5}(f).
 These considerations apply, of course, only when the material is flowing down. In a static bed, a large part of the weight would be counterbalanced by the HGS material below, upon which the ball rests (as in the hydrostatic analogue). Then, the ball would merely experience a reduction of its weight by buoyancy.
%
In agreement with this, we find a sharp drop of $F$ when the outlet is closed. The contribution of friction ceases immediately. A subsequent further reduction of the measured force indicates that the HGS compactify below the ball until they counterbalance the weight forces from above, analogous to buoyancy in ordinary liquids.

Some correction is yet necessary to this simple model: Soft HGS in static beds have larger fill fractions than expected from a random close sphere packing, because they deform. In narrow containers, the difference between the static packing fraction and that during discharge was found to be between 5 \% and 15 \%, increasing with depths $h$ \cite{Harth2020}. Thus, the fill height is not exactly linear with the discharged mass. One has to take into account that the packing first dilates by roughly 10~\% before the bed height sinks.

In the hard-grain silo, the situation is qualitatively different. Figure~\ref{fig:sketch5}(e) represents the force scaled with the obstacle radius (see numerical in the inset). Evidently, the dependence is nearly linear.
Our force measurements are in reasonable agreement with earlier observations \cite{Wieghardt1952,Wieghardt1975,Roth2021,Chehata2003} that the  velocity of an intruder relative to the flowing material has almost no influence on the drag forces
(Fig.~\ref{fig:forceflow}). At least they are much weaker than in the HGS material. The remaining small effect of flow in our experiment is apparently a temporary disruption of force chains that redirect some part of the weight of the upper material onto the obstacle.

An estimate of the pressures in the GLS silo can be obtained from force measurements at the bottom plate \cite{Supplement}. Before the discharge starts, 80~N are measured. During the discharge, the force lowers to a plateau of  $\approx 40$~N, corresponding to a mean pressure of 2460 Pa, or the weight of a 16~cm  GLS layer. This plateau ends
when the fill level drops below $\approx 24$~cm and the force gradually decreases to zero.
From these data, one may roughly estimate the force generated by the material on top of the obstacle. If one considers only the horizontal cross-section of the ball, as in the HGS case, one finds an equivalent of slightly more than 3 N for the 40 mm ball, and only 0.2 N for the 10 mm ball. The actual forces are substantially larger and much less ball-size dependent, which leads to the conclusion that the obstacle carries a considerable amount of the weight of grains that are not directly above its cross-section.  This explains the much lower radius dependence of the measured forces.
As the pressure conditions should remain nearly constant until $h$ becomes comparable to $D_{\rm silo}$,
the forces related to the weight of overlying grains are expected to remain nearly unchanged until 45~\% 
of the material is discharged
(fill level lowered by $\approx 36$~cm). This is quite well reproduced in Fig. \ref{fig:sketch5} at least for the higher discharge rates.

\begin{figure}
\centering
\includegraphics[width=\columnwidth]
{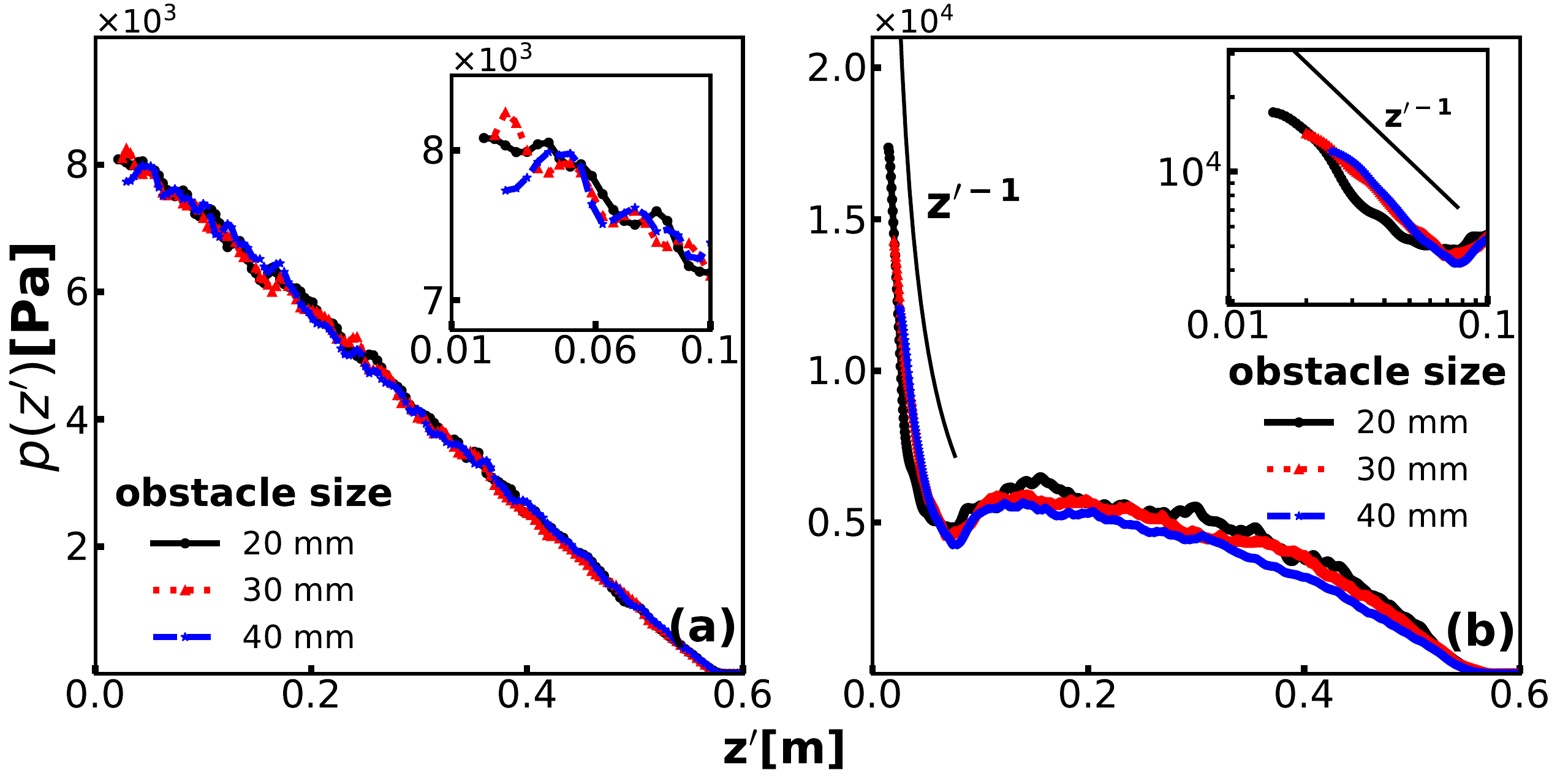}
\includegraphics[width=\columnwidth]
{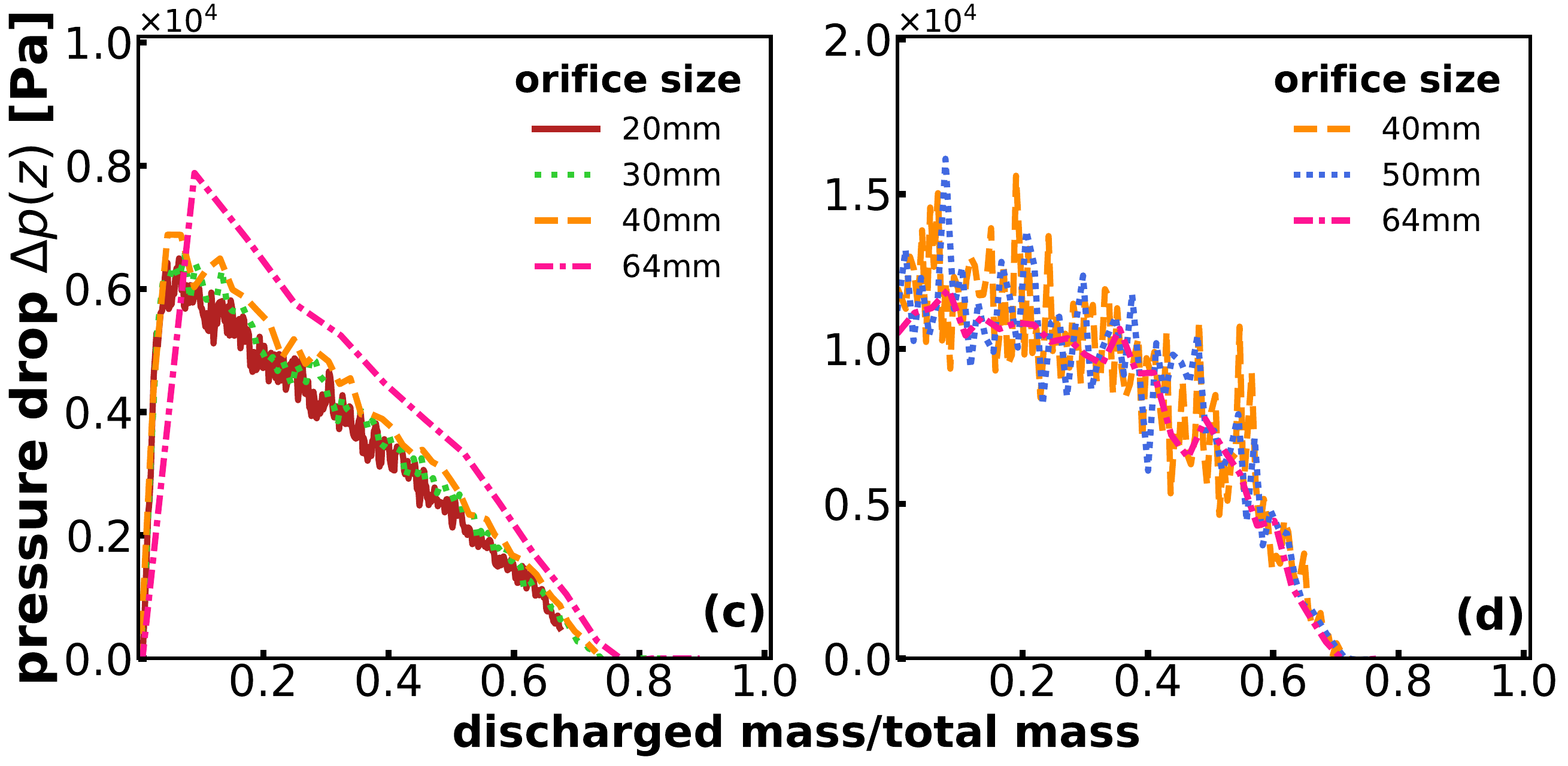}
\caption{The pressure profiles $p(z)$ obtained numerically above the intruder for HGS (a) and GLS (b), silo with an orifice size $D_0 = 64$~mm,
the insets show zoom of the profiles near the obstacle (log-log in panel (b)).
The mean pressure drop (obstacle size $d=40$~mm), depending on the outflow mass in terms of the total mass, for HGS (c) and GLS (d).
}
\label{fig:press_drop}
\end{figure}

The obstacle size dependence of the forces remains to be understood. For that purpose, we performed a micromechanical analysis of the numerical results.
Post-processing the DEM data, we computed the macroscopic stress tensor $\sigma(z)$ above the obstacle, averaged in a cylindrical region with the obstacle's lateral dimensions.
In the following analysis, we focus on the pressure profile in the $z$ direction, especially above the obstacle center $z_{\rm obst}$, ($z'=z-z_{obst}$). The macroscopic pressure is defined as the trace of the stress tensor $p(z)=1/3 \cdot Tr(\sigma(z))$.

Figures~\ref{fig:press_drop}(a,b) exemplify the pressure profiles $p(z')$ obtained numerically for HGS (Fig.~\ref{fig:press_drop}(a))  and GLS (Fig.~\ref{fig:press_drop}(b)).
The insets show data obtained at distances of the order of the obstacle diameter.
The graphs include results corresponding to several obstacle sizes at specific times but with the same orifice size $D_0 = 64$ mm.
In the case of the low-friction soft particles (Fig.~\ref{fig:press_drop}(a)), the pressure increases linearly
with depth, being highest at the obstacle's surface.
The inset of Fig.~\ref{fig:press_drop}(a) shows that the variations of $p(z')$ within a distance comparable with the obstacle sizes are irrelevant.
Consequently, the total force exerted on the obstacle surface in each case is expected to be proportional to the obstacle area $F \propto A_o$, in excellent agreement with the data in Fig.~\ref{fig:sketch5}(e).

In the case of hard particles, however, the picture is entirely different (Fig.~\ref{fig:press_drop}(b)).
Starting from the top of the granular bed, the pressure increases until reaching a "plateau", where changes in pressure are very low. Interestingly, close to the obstacle, the pressure abruptly increases
reflecting a spatial variation $p(z') \sim 1/z'$ (see also the inset).
The data conclusively indicate that the pressure at the obstacle surface roughly scales with the size as $p \sim 1/R$ (note the leftmost points of the curves in Fig.~\ref{fig:press_drop}(b)). These conditions remain unaltered during a large part of the discharge process. Accordingly, the pressure projection and integration on the obstacle surface results in a total force that scales as $F \sim R$, explaining %
the findings shown in Fig.~\ref{fig:sketch5}(f).

Figures~\ref{fig:press_drop}(c,d) show the mean pressure drop $\langle \Delta p(z) \rangle$ across the obstacle as a function of the mass remaining in the silo, obtained for an obstacle diameter $d = 40$ mm, in HGS and GLS, respectively.
The graphs include data
corresponding to several orifice sizes. In Fig.~\ref{fig:press_drop}(c), the data indicate that 
low-friction soft particles are exposed to a pressure drop that significantly varies as the discharge process goes on. Besides,
the discharge flow rate (orifice size) notably impacts the $\langle \Delta p(z) \rangle$ values.
In contrast, for GLS the pressure gradient is practically the same for all the explored orifices, and it does not vary during a significant part of the discharge process (see Fig.~\ref{fig:press_drop}(d)).
In both materials, the computed pressure drops strongly correlate with the forces acting on the obstacles as illustrated in Figs.~\ref{fig:sketch5}(a) and ~\ref{fig:sketch5}(b), respectively.


Concluding, we have shown that a sphere suspended in a discharging silo experiences mechanical forces from the weight of the overlaying layers and the friction of the surrounding moving granular material.
In experiments and simulations with hard frictional glass particles, the force on the obstacle was nearly uninfluenced by the flow velocity.
Its value remained unaltered during a large part of the discharge process and depended linearly on the obstacle diameter.
The simulations indicate that during the discharge, the pressure of the granular bed at the obstacle's surface scales with the size as $p \sim 1/R$, which is congruent with a force proportional to the diameter of the obstacle. Besides, the mean pressure gradient acting on the obstacles was practically the same for all the explored orifices and did not vary significantly in the discharge process. It is worth mentioning that
when the outflow of GLS was interrupted,  the force on the ball remained nearly unchanged, indicating the predominantly static nature of the interaction.
On the other hand, in flowing frictionless soft particles, noticeable drag is added to the gravitational forces on the suspended obstacles.
As confirmed by our micromechanical analysis, the obstacle experienced a total force from the top as if immersed in a dynamic hydrostatic pressure profile, but practically without acting from below.
Irrespective of the low friction, the particle collisions generate a noticeable drag force. It increases with velocity up to a certain speed, but then reaches saturation. We assume that at high discharge rates, the local packing density drops, thus reducing the drag force and compensating for the effects of increased velocity.

In addition to the new insights in forces on obstacles in flowing granular material, the present experiments and DEM simulations might also provide an interesting geometry to test nonlocal flow models \cite{Kamrin2012,Henann2013}.

This project received funding from the European Union's Horizon 2020 Research and Innovation program under the Marie Sk\l{}odowska-Curie grant agreement, CALIPER No. 812638. RCH acknowledges the Ministerio de Ciencia e Innovaci\'on (Spanish Government) Grant PID2020-114839GB-I00 funded by MCIN/AEI/10.13039/501100011033. TP acknowledges support by the Startup Grant from DKU. The Collective Dynamics Lab is partly sponsored by a philanthropic gift.

%

\end{document}